\documentclass[aps, prl, twocolumn, 10pt, superscriptaddress, nofootinbib]{revtex4-2}
\usepackage[dvipsnames]{xcolor}
\usepackage{physics}
\usepackage{braket}
\usepackage{scalerel}
\usepackage{blindtext} 
\usepackage{epsfig, cancel}

\usepackage{latexsym}
\usepackage{natbib, comment}
\usepackage{mathrsfs,amsmath,amssymb,amsthm,amsfonts,tikz,graphicx,accents,hyperref,color}
\usepackage{url}
\usepackage{dcolumn}
\usepackage{multirow}
\usepackage{color}
\usepackage{cancel}
\usepackage{soul}
\usepackage[normalem]{ulem}
\usepackage{txfonts}
\usepackage{epsfig}
\usepackage{psfrag}
\usepackage{subfigure}
\hypersetup{colorlinks=true}
\usepackage{mathtools}
\usepackage{enumitem}
\usepackage{float}
\usepackage{caption,ragged2e}
\usetikzlibrary{decorations.markings,fpu}

\usetikzlibrary{calc}

\def\H0{{\text{H}\hspace*{-2.05mm}\text{H} 0\hspace*{-1.35mm}0\ }}

\usepackage[all]{xy}

\usepackage{slashed}
\usepackage{slashed,ccaption}
\usepackage{multirow}
\usetikzlibrary{calc} 

\usepackage{caption}

\newcommand{\beps}{\boldsymbol{\epsilon}}
\usepackage{array}
\hypersetup{ linktoc=all,
    colorlinks, linkcolor={palatinateblue},
    citecolor={red}, urlcolor={amaranth} %{brightpink}
}

\graphicspath{{Images/}}

\renewcommand{\d}[1]{\ensuremath{\operatorname{d}\!{#1}}}

\DeclareSymbolFont{extraup}{U}{zavm}{m}{n}
\DeclareMathSymbol{\varheart}{\mathalpha}{extraup}{86}
\DeclareMathSymbol{\vardiamond}{\mathalpha}{extraup}{87}
\makeatletter
\renewcommand*{\@fnsymbol}[1]{\ensuremath{\ifcase#1\or \clubsuit \or \vardiamond \or \varheart\or
    \spadesuit\or \mathparagraph\or \|\or **\or \dagger\dagger
    \or \ddagger\ddagger \else\@ctrerr\fi}}
\makeatother

\newcommand{\observer}{\tikz[baseline=-0.3ex, x=0.75ex, y=0.75ex, line width=0.12ex, line cap=round]{
  \draw (0,1.2) circle (0.5);                % Head
  \draw (0,0.7) -- (0,-0.3);                 % Torso
  \draw (-0.6,0.2) -- (0,0.5) -- (0.6,0.2);  % Arms
  \draw (-0.6,-0.9) -- (0,-0.3) -- (0.6,-0.9); % Legs
}}

\definecolor{rosy}{RGB}{230,235,252}
\definecolor{myframetitle}{RGB}{90,89,170}
\definecolor{myblocktitle}{RGB}{140,185,249}
\definecolor{mytitle}{RGB}{10,80,26}

\definecolor{darkgreen}{RGB}{27,130,45}
\definecolor{darkblue}{rgb}{0,0,0.3}
\definecolor{darkred}{rgb}{0.7,0,0}

\definecolor{light gray}{RGB}{220,220,220}
\definecolor{dark purple}{RGB}{108,0,217}
\definecolor{pink}{RGB}{190,20,100}
\definecolor{orang}{RGB}{193,63,0}
\definecolor{green}{RGB}{11,98,17}
\definecolor{darkpink}{RGB}{153,0,76}
\definecolor{bluegreen}{RGB}{0,102,102}
\definecolor{greenlagan}{RGB}{0,102,0}
\definecolor{redgreen}{RGB}{102,102,0}
\definecolor{Redgreen}{RGB}{153,76,0}
\definecolor{vividviolet}{rgb}{0.62, 0.0, 1.0}
\definecolor{amaranth}{rgb}{0.9, 0.17, 0.31}
\definecolor{palatinateblue}{rgb}{0.15, 0.23, 0.89}
\definecolor{brightpink}{rgb}{1.0, 0.0, 0.5}
\definecolor{cornflowerblue}{rgb}{0.39, 0.58, 0.93}
\definecolor{deepcarminepink}{rgb}{0.94, 0.19, 0.22}
\definecolor{radicalred}{rgb}{1.0, 0.21, 0.37}
\newcommand\ignore[1]{}
\usepackage[most]{tcolorbox}

\tcbset{highlight math style={left=02mm,right=02mm,top=02mm,bottom=02mm}} %MHV make box for equations
\usepackage{empheq}

\newcommand\inbox[1]{\tcbset{fonttitle=\scriptsize} \tcboxmath[colback=white,colframe=black!70]{#1}}
\begin{document}

\title{Who Writes the Gravitational Second Law of Thermodynamics?}

\author{V.~R.~Shajiee}\email{v.shajiee@ipm.ir}
\affiliation{School of Physics, Institute for Research in Fundamental Sciences (IPM), P.O.Box 19395-5531, Tehran, Iran}

\author{M.~M.~Sheikh-Jabbari}\email{jabbari@theory.ipm.ac.ir}
\affiliation{School of Physics, Institute for Research in Fundamental Sciences (IPM), P.O.Box 19395-5531, Tehran, Iran}

\author{V.~Taghiloo}\email{v.taghiloo@iasbs.ac.ir}
\affiliation{Department of Physics, Institute for Advanced Studies in Basic Sciences (IASBS), P.O.Box 45137-66731, Zanjan, Iran}
\affiliation{School of Physics, Institute for Research in Fundamental Sciences (IPM), P.O.Box 19395-5531, Tehran, Iran}

\begin{abstract}
We define gravitational entropy as the manifestly integrable surface charge associated with local transverse Lorentz boosts, entirely bypassing the conventional reliance on spacetime diffeomorphisms. Any notion of entropy must satisfy the second law. To answer the question posed in the title, an analogy with Newtonian classical mechanics is instructive: Newton's second law of motion is written by inertial observers who are defined by the first law of mechanics. We show that the second law of gravitational thermodynamics is written by free-fall observers with path parameterization in which the non-affinity of the geodesics is equal to the expansion of their velocity vector field. We then provide a proof of the local second law by studying variations in entropy as viewed by this class of covariantly-defined causal free-fall observers. We show that the entropy variation is strictly non-decreasing, provided the matter sector satisfies the integrated strong energy condition along the observer path.
\end{abstract}
\maketitle
%%%%%%%%%%%%%%%%%%%%%%%%%%%%%%%%%%%%%%%%%%%%%%%%%%%%%%%%%%%%%%%%%%%%%%
%\section{Introduction}
%%%%%%%%%%%%%%%%%%%%%%%%%%%%%%%%%%%%%%%%%%%%%%%%%%%%%%%%%%%%%%%%%%%%%%%%

The relation between gravity and thermodynamics is profound, deep, old, and still an active research area; see e.g., \cite{Grumiller:2020vvv} for historical reviews. It started in the early 1970s with studies involving black holes \cite{Bekenstein:1972tm, Bekenstein:1973ur, Hawking:1973uf}. In the early 1990s, two important developments occurred: (1) Wald provided a concise definition of the entropy as a conserved Noether charge for generic diffeomorphism-invariant theories of gravity for stationary black holes with a bifurcate Killing horizon \cite{Wald:1993nt, Iyer:1994ys}; (2) Jacobson provided a derivation of Einstein's equations from the Clausius relation, in a setup that involves null surfaces \cite{Jacobson:1995ab}. Taking into account Jacobson's viewpoint motivates one to extend and generalize Wald's mathematically robust formulation beyond stationary black holes, to cases that are closer to realistic objects one finds in the sky, and also to cosmological settings. See e.g. \cite{Jacobson:1993pf, Jacobson:1993vj, Hayward:1993wb, Ashtekar:2002ag, Ashtekar:2003hk, Andersson:2005gq, Booth:2005qc, Andersson:2007gy, Bousso:1999xy, Wall:2009wm, Wall:2011hj, Hollands:2022fkn, Hollands:2024vbe, Shajiee:2026coz}, for an incomplete list of references.

Among thermodynamic notions, entropy has a very special place, and its evolution, which is subject to the second law of thermodynamics, dictates how thermodynamic systems evolve. From the statistical mechanics viewpoint, entropy is related to the logarithm of microscopic degrees of freedom and, as such, it is observer-independent. However, its variations, in principle, can depend on the observer who records the changes, through the local choice of time units. Within the gravitational context, this point becomes even more sensitive and crucial, as the defining principle of general relativity is that no reference frame is preferred: the laws of gravity are formulated covariantly and apply to all admissible observers. 

The notion of covariance should be explored in view of the thermodynamical description of gravitational systems. Unruh took a crucial step in this direction, extending Einstein's equivalence principle to the semiclassical regime. He showed that an accelerated observer on flat Minkowski space locally records a thermal physics description \cite{Unruh:1976db}, while a non-accelerating observer provides a standard non-thermal physics description. Similarly, a free-fall observer passing through the horizon of a black hole locally records a non-thermal description. So, the thermal description of physical systems within general relativity may be observer-dependent. This 
naturally raises a similar fundamental question about the entropy and the second law: \textit{Who writes the gravitational second law of thermodynamics?}

In this Letter, we depart from this standard paradigm to answer the question posed above. Unlike previous approaches which rely on spacetime diffeomorphisms, we define gravitational entropy as the surface charge associated with local boosts transverse to codimension-2 spacelike surfaces, as part of local Lorentz symmetries \cite{Jacobson:2015uqa, Godazgar:2020kqd, Shajiee:2025cxl}. This allows us (1) to detach the definition of the charge from codimension-1 boundaries (e.g., horizons or null surfaces); (2) to formulate the role of observers in the gravitational second law for generic observers; and (3) to study the second law as viewed by generic \textit{causal} observers, rather than commonly considered null surfaces/observers. We demonstrate that the gravitational second law is written precisely by a special class of locally free-falling causal observers, equipped with a path parametrization adjusted by the expansion of the velocity vector field of the geodesic. For this rigorously defined class of observers, we prove that the entropy variation is strictly non-decreasing, provided the spacetime matter content satisfies the integrated strong energy condition along their trajectory.

\begin{center}
%%%%%%%%%%%%%%%%%%%%%%%%%%%%%%%%%%%%%%%%%
\textbf{Geometric setup and kinematics}\label{sec:II}
%%%%%%%%%%%%%%%%%%%%%%%%%%%%%%%%%%%%%%%%%
\end{center}

\begin{figure}[htb]
\centering
\begin{tikzpicture}[>=stealth, scale=1.7]
  % 1. Shaded Future and Past Lightcones
  \fill[gray!10] (0,0) -- (1.4,1.4) -- (-1.4,1.4) -- cycle;
  \fill[gray!10] (0,0) -- (1.4,-1.4) -- (-1.4,-1.4) -- cycle;

  % Lightcone boundaries (dashed)
  \draw[dashed, gray!80] (-1.4,-1.4) -- (1.4,1.4);
  \draw[dashed, gray!80] (-1.4,1.4) -- (1.4,-1.4);

  % 2. Spacetime Axes
  \draw[->, thick, gray!70!black] (-1.7,0) -- (1.7,0) node[right, text=black] {$r$};
  \draw[->, thick, gray!70!black] (0,-1.7) -- (0,1.7) node[above, text=black] {$\tau$};

  % 3. Causal curve \gamma (Strictly concave, always expanding, always slowing down)
  % r(t) = 0.5*t - 0.2917*ln(0.5*(exp(1.2*t) + exp(-1.2*t)))
  \draw[very thick, blue!80!black, decoration={markings, mark=at position 0.85 with {\arrow{>}}}, postaction={decorate}]
    plot[domain=-1.4:1.4, samples=100, variable=\t] 
    ({0.5*\t - 0.2917*ln(0.5*(exp(1.2*\t) + exp(-1.2*\t)))}, {\t})
    node[right] {\large $\gamma$};

  % 4. Right-angle marker (proving Euclidean orthogonality between s and v at slope = 2.0)
  \draw[black!60, thick] (-0.13, 0.07) -- (-0.06, 0.20) -- (0.07, 0.13);

  % 5. Tangent vector v and Normal vector s (orthogonal in the 2D plane at t=0)
  \draw[->, very thick, cyan!80!black] (0,0) -- (0.45, 0.90)  node[above right=-2pt, inner sep=1pt, fill=white, fill opacity=0.7, text opacity=1] {$v^{\mu}_{\lambda}$};
  \draw[->, very thick, green!60!black] (0,0) -- (-0.90, 0.45) node[left] {$s^\mu$};

  % 6. Point representing the codimension-2 surface Sigma
  \filldraw[black] (0,0) circle (1.5pt);
  \node[below right=2pt, font=\bfseries] at (0,0) {$\Sigma(\lambda)$};

  % 7. Plane \mathcal{D} symbol (Top Right)
  \begin{scope}[shift={(1.3, 1.4)}]
    \node at (0.2, 0.3) {$\mathcal{D}$};
    \draw[->, thick] (0,0) -- (0.4,0);
    \draw[->, thick] (0,0) -- (0,0.4);
  \end{scope}
\end{tikzpicture}
\caption{\justifying
Future-oriented causal trajectory of the observer $\gamma$ in the $2d$ plane $\mathcal{D}$. The local frame is adapted such that the tangent vector $v^{\mu}_{\lambda}$ and the spacelike normal vector $s^\mu$ are orthogonal. As $\lambda \to \infty$, the curve asymptotes to a constant radius, representing the expansion $\theta_{v_\lambda}$ vanishing equilibrium state.}
\label{fig:setup-1}
\end{figure}
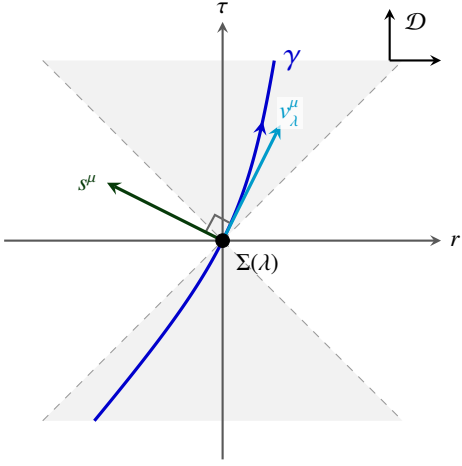

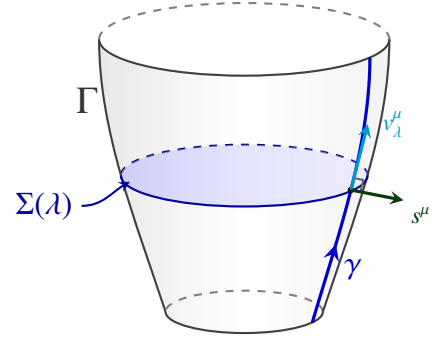
\begin{figure}[t]
\centering
\begin{tikzpicture}[scale=.8, >=stealth,
    surface/.style={left color=gray!30, right color=gray!15, middle color=white, opacity=0.6},
    boundary/.style={thick, draw=black!75},
    hidden/.style={dashed, draw=black!50, thick},
    slicefill/.style={left color=blue!30, right color=blue!10, opacity=0.5}
  ]

  % Core dimensions for the causal tube
  \def\H{4.5}                  
  \def\Rb{1.3} \def\ryb{0.35}  
  \def\Rt{2.4} \def\ryt{0.6}   

  % Intermediate slice \Sigma(\lambda) parameters (Evaluated at bezier t=0.5)
  \def\Rm{2.04}   
  \def\rym{0.51} 
  \def\Ym{2.25}   

  % 1. Shaded surface for the causal boundary \Gamma (Expanding then reaching equilibrium)
  \shade[surface]
    (-\Rb, 0) .. controls (-1.8, 1.5) and (-2.4, 3.0) .. (-\Rt, \H) 
    arc [start angle=180, end angle=360, x radius=\Rt, y radius=\ryt]
    .. controls (2.4, 3.0) and (1.8, 1.5) .. (\Rb, 0) 
    arc [start angle=360, end angle=180, x radius=\Rb, y radius=\ryb];

  % 2. Hidden (back) boundaries
  \draw[hidden] (\Rb, 0) arc [start angle=0, end angle=180, x radius=\Rb, y radius=\ryb];
  \draw[hidden, draw=blue!50!black] (\Rm, \Ym) arc [start angle=0, end angle=180, x radius=\Rm, y radius=\rym];
  \draw[hidden] (\Rt, \H) arc [start angle=0, end angle=180, x radius=\Rt, y radius=\ryt];

  % 3. Highlighted spacelike cross-section \Sigma(\lambda)
  \shade[slicefill] (0, \Ym) ellipse [x radius=\Rm, y radius=\rym];

  % 4. Outer side walls (Bezier curves showing asymptotic equilibrium)
  \draw[boundary] (-\Rb, 0) .. controls (-1.8, 1.5) and (-2.4, 3.0) .. (-\Rt, \H);
  \draw[boundary] (\Rb, 0) .. controls (1.8, 1.5) and (2.4, 3.0) .. (\Rt, \H);

  % 5. Visible (front) boundaries
  \draw[boundary] (-\Rb, 0) arc [start angle=180, end angle=360, x radius=\Rb, y radius=\ryb];
  \draw[boundary, draw=blue!60!black] (-\Rm, \Ym) arc [start angle=180, end angle=360, x radius=\Rm, y radius=\rym];
  \draw[boundary] (-\Rt, \H) arc [start angle=180, end angle=360, x radius=\Rt, y radius=\ryt];

  % --- Annotations and Vectors ---
  
  \node[font=\Large\bfseries, text=black!80] at (-2.6, 3.5) {$\Gamma$};

  % Label for the spacelike boundary \Sigma(\lambda)
  \coordinate (SigmaPointer) at ({-\Rm + 0.15}, {\Ym - 0.1});
  \draw[<-, thick, blue!60!black] (SigmaPointer) to[out=210, in=0] ++(-0.8, -0.4) node[left] {\large $\Sigma(\lambda)$};

  % Draw the curve \gamma on the surface with the arrow moved to position 0.3
  \draw[very thick, blue!80!black, decoration={markings, mark=at position 0.3 with {\arrow[scale=1.2]{>}}}, postaction={decorate}]
    plot[domain=0:1, samples=40, variable=\t]
    ( { (1.3*(1-\t)*(1-\t)*(1-\t) + 5.4*(1-\t)*(1-\t)*\t + 7.2*(1-\t)*\t*\t + 2.4*\t*\t*\t)*0.866 },
      { 4.375*\t - 0.175 }
    );

  % Repositioned label for \gamma (placed clearly outside the shaded boundary)
  \node[blue!80!black] at (1.8, 0.7) {\large $\gamma$};

  % Specific evaluation point on \Sigma(\lambda)
  \coordinate (Gmid) at (1.764, 2.012);
  \fill[black] (Gmid) circle (1.5pt);

  % Vector definitions for tangent and normal calculated exactly from the surface gradient
  \def\vx{0.28} \def\vy{1.09} 
  \def\sx{0.87} \def\sy{-0.17} 

  \pgfmathsetmacro{\mkvx}{0.18 * \vx}
  \pgfmathsetmacro{\mkvy}{0.18 * \vy}
  \pgfmathsetmacro{\mksx}{0.18 * \sx}
  \pgfmathsetmacro{\mksy}{0.18 * \sy}
  
  \draw[thick, black!60, line cap=round, line join=round]
    ($(Gmid) + (\mkvx, \mkvy)$) --
    ($(Gmid) + (\mkvx + \mksx, \mkvy + \mksy)$) --
    ($(Gmid) + (\mksx, \mksy)$);

  % Tangent vector v^{\mu}
  \draw[->, very thick, cyan!80!black] (Gmid) -- ++(\vx, \vy) node[right=2pt, font=\bfseries] {$v^{\mu}_{\lambda}$};

  % Normal vector s^\mu
  \draw[->, very thick, green!60!black] (Gmid) -- ++(\sx, \sy) node[below right=-1pt, font=\bfseries] {$s^\mu$};

\end{tikzpicture}
\caption{\justifying
The codimension-1 causal surface $\Gamma$ which is topologically $\Sigma \times \gamma$. At a given $\lambda$, $\Gamma$ is limited to the spacelike boundary $\Sigma(\lambda)$. As $\lambda \to \infty$, the surface reaches an equilibrium state where the expansion vanishes ($\theta_{v_\lambda} \to 0$).}
\label{fig:setup-2}
\end{figure}

Consider a $D$-dimensional spacetime containing a compact, finite-area, spacelike codimension-2 surface $\Sigma$. In our framework, the gravitational entropy is defined as an integral over this surface. This entropy is measured by an observer traveling along a future-directed timelike curve $\gamma$ normal to $\Sigma$ (see Fig.~\ref{fig:setup-1}). Together, $\Sigma$ and $\gamma$ span a codimension-1 causal hypersurface, locally and topologically $\Gamma \simeq \Sigma \times \gamma$ (see Fig.~\ref{fig:setup-2}).

Let $\mathcal{D}$ denote the 2-dimensional normal bundle to $\Sigma$. An independent observer's trajectory $\gamma$ lies within $\mathcal{D}$, which is spanned by their future-directed timelike velocity $v^\mu$ (tangent to $\gamma$) and an orthogonal spacelike normal vector $s^\mu$:
\begin{equation}\label{v-s-frame}
   v\cdot v=-1\,,\qquad s\cdot s=1\,,\qquad v\cdot s=0\,.
\end{equation}
By adapting the local frame to this observer and denoting the induced metric on $\Sigma$ by $q_{\mu\nu}$, the full background spacetime metric $g_{\mu\nu}$ decomposes as
\begin{equation}\label{g-q-transerve}
   g_{\mu\nu}= q_{\mu\nu}- v_\mu v_\nu + s_\mu s_\nu\,, \qquad  q_{\mu\nu} v^\nu=0=q_{\mu\nu} s^\nu\,.
\end{equation}
The binormal to $\Sigma$ is then defined as
\begin{equation}\label{binormal}
    {\beps}_{\mu\nu}:= -v_\mu s_\nu + s_\mu v_\nu\,,\qquad  {\beps}_{\mu\nu} {\beps}^{\mu\nu} =-2\,.
\end{equation}

The covariant derivatives of the basis vectors $v^\mu$ and $s^\mu$ generally have components outside $\mathcal{D}$ and can be parametrically decomposed as
\begin{subequations}\label{derv-v-s}\begin{align}
     \nabla_{\mu}v_{\nu} &=  \kappa_v v_\mu s_\nu  - \kappa_s s_\mu s_\nu - v_\mu a^{(v)}_\nu - s_\mu \eta^{(v)}_\nu -\omega_\mu s_\nu + \Theta^{(v)}_{\mu\nu} \,,\label{CD-v-inde}\\
      \nabla_{\mu}s_{\nu} &=  \kappa_v v_\mu v_\nu - \kappa_s s_\mu v_\nu + s_\mu a^{(s)}_\nu + v_\mu \eta^{(s)}_\nu - \omega_\mu v_\nu + \Theta^{(s)}_{\mu\nu} \, , 
      \label{CD-s-inde}
\end{align}\end{subequations}
where $\Theta^{(v,s)}_{\mu\nu}$ are the spatial deviation tensors (extrinsic curvatures) on $\Sigma$, $\kappa_{v,s}$ dictate the cross-accelerations in the transverse plane, $\omega_\mu$ is the rotational 1-form, and $a^{(v,s)}_\mu, \eta^{(v,s)}_\mu$ are the transverse connection 1-forms. 

\textit{Kinematic choices.} The above equations define the derivatives of $v^\mu$ and $s^\mu$ locally, leaving the freedom to constrain their transport along $\Gamma$. Let $\mathrel{\hat{=}}$ denote an equality evaluated strictly on $\Gamma$. We first assume that $v^{\mu}$ is hypersurface orthogonal on $\Gamma$. This eliminates its twist and enforces
\begin{equation}\label{hyper-surface-ortho-1}
    \eta^{(v)}_\mu  \mathrel{\hat{=}} \omega_\mu\, .
\end{equation}
Being twist-free implies that $\Theta^{(v)}_{\mu\nu}$ is symmetric, allowing it to be decomposed solely into its trace (the expansion $\theta_v$) and its symmetric traceless part (the shear $N^{(v)}_{\mu\nu}$):
\begin{equation}\label{deviation-tensors-decomposed}
     \Theta^{(v)}_{\mu\nu} \mathrel{\hat{=}} \frac{\theta_{v}}{D-2}q_{\mu\nu}+N^{(v)}_{\mu\nu}\,.
\end{equation}
Furthermore, we utilize the gauge freedom in the normal direction to set
\begin{equation}\label{s-affine}
    \kappa_s \mathrel{\hat{=}} 0\, .
\end{equation}
The remaining degrees of freedom, such as the path parameterization and the transverse acceleration $a^{(v)}_\mu$, correspond to the physical choices of the observer. These will be strictly fixed later by demanding that the observer follows a causal geodesic with a specific path parameterization.

\begin{center}
%%%%%%%%%%%%%%%%%%%%%%%%%%%%%%%%%%%%%%%%%%%%%%%%%%%%%%%%%%
\textbf{{Entropy is the charge associated with the local boosts}}\label{sec:II-b}
%%%%%%%%%%%%%%%%%%%%%%%%%%%%%%%%%%%%%%%%%%%%%%%%%%%%%%%%%%
\end{center}

We define the gravitational entropy of a region of space enclosed by a compact, finite-area codimension-2 surface $\Sigma$ as the conserved surface charge associated with local Lorentz boosts in the transverse plane ${\cal D}$. To evaluate this charge, it is most natural to employ the first-order (tetrad) formulation of gravity. In this framework, the dynamical fields are the frame field 1-forms $e^a$ and the spin connection 1-forms $\omega^{ab}$, making local Lorentz transformations manifest gauge symmetries. We compute the charge within the covariant phase space formalism (CPSF) \cite{Lee:1990nz, Barnich:2007bf, Harlow:2014yka, Compere:2018aar, Grumiller:2022qhx}. The details of the analysis may be found in \cite{Shajiee:2025cxl}. This entropy for a generic theory of gravity, when $\Sigma$ is a bifurcate Killing horizon, reproduces the Wald entropy \cite{Wald:1993nt}. Hereafter, and to make the analysis of the second law more tractable, we restrict ourselves to Einstein gravity theory. 

Let $\Lambda^{ab}$ denote the local boost symmetry generator whose charge is the entropy. In Einstein gravity, the associated charge variation is given by \cite{Jacobson:2015uqa, Godazgar:2020kqd}
\begin{equation}\label{charge-var-in-D}
    \delta Q_{\Lambda} 
    = -\frac{1}{16\pi G}\int_{\Sigma}\Lambda^{ab} \delta \big(\star(e_{a}\wedge e_{b})\big) \, ,
\end{equation}
where $\star$ is the Hodge dual. Assuming the symmetry parameter $\Lambda^{ab}$ to be independent of the dynamical fields ($\delta\Lambda^{ab}=0$), this charge variation is manifestly integrable, yielding 
\begin{equation}\label{entropy-general-def}
Q_{\Lambda} = -\frac{1}{16\pi G}\int_{\Sigma} \star(e_{a}\wedge e_{b})\,\Lambda^{ab}\, .
\end{equation}

In the notation adapted in the geometric setup part, the local boost generator in the ${\cal D}$ plane is proportional to the binormal to $\Sigma$ with a universal proportionality constant factor:
\begin{equation}\label{local-boost-generator}
   \Lambda^{ab}_{\text{boost}}=-\frac{2\pi}{\hbar}\beps^{ab}\,,
\end{equation}
where $\beps^{ab}$ is the frame-basis equivalent of the binormal $\beps_{\mu\nu}$ defined in \eqref{binormal}, $\beps^{ab}=\beps^{\mu\nu}e_\mu^a e^b_\nu$. Substituting this symmetry generator into the integrated charge immediately yields the standard ``area law'' for the gravitational entropy,
\begin{equation}\label{entropy-def-EH}
    S[\lambda] = \frac{1}{4 G \hbar} \int_{\Sigma(\lambda)} \d{}^{D-2}x\, \sqrt{q} = \frac{A(\lambda)}{4G\hbar}\, .
\end{equation}
%\tcr{Related connections between area and surface boosts were developed by Donnelly and Freidel, while internal Lorentz corner charges in tetrad gravity were analyzed by Freidel, Geiller and Pranzetti \cite{Freidel:2020xyx,Donnelly:2016auv}. Here we identify gravitational entropy with the specifically normalized transverse Lorentz charge in Eqs. \eqref{entropy-general-def}–\eqref{entropy-def-EH} and investigate its evolution along causal observers.}
This expression is to be evaluated on a spacetime satisfying the Einstein field equations. Although this formulation of entropy shares mathematical similarities with the constructions of Wald, Jacobson–Kang–Myers, and Bousso \cite{Wald:1993nt, Jacobson:1993vj, Bousso:1999xy, Bousso:2002ju}, its physical interpretation is distinctly different. A detailed comparison with other definitions of gravitational entropy may be found in \cite{Shajiee:2025cxl}, here we only emphasize the fact that the surface charge we have identified as the entropy is manifestly integrable over the solution space and entirely bypasses the need for a field-dependent surface gravity normalization, which often introduces non-integrability and observer-dependence issues in standard Killing-vector-based formulations \cite{Iyer:1994ys, Hajian:2015xlp, Hajian:2020dcq}.

\begin{center}
%%%%%%%%%%%%%%%%%%%%%%%%%%%%%%%%%%%%%%%%%%%%%%%%%%%%%%%%%%%%%%%%%%%%%%%%%%%%%%%%%%
\textbf{Who writes the second law?}\label{sec:observer}
%%%%%%%%%%%%%%%%%%%%%%%%%%%%%%%%%%%%%%%%%%%%%%%%%%%%%%%%%%%%%%%%%%%%%%%%%%%%%%%%%%
\end{center}

Having defined the entropy, the next immediate analysis is to examine the second law. The second law is about how a physical observer measures variations in the entropy as time advances. Given the definition of the entropy above, as the charge associated with local boosts and that this definition involves no symmetry generator \textit{vector field} (in contrast, e.g. to Wald's entropy or its more modern reincarnations as ``dynamical entropy'' \cite{Hollands:2024vbe, Shajiee:2026coz}),\footnote{From a vector field one can infer an observer, e.g. viewing the vector field as the velocity vector associated with an observer, if the vector field is causal. If it is not causal, one can consider the vector field normal to it in the $2d$ plane ${\cal D}$. In our case, however, we are dealing with a 2-form as a symmetry generator.} one is led to the question \textit{Who formulates the gravitational second law?}

To answer the question posed above, we construct the most general parameterization of a causal curve in the plane $\mathcal{D}$. Consider a reference observer following a causal trajectory $x^\mu=x^\mu(\tau)$, parameterized by proper time $\tau$. The corresponding future-directed causal velocity field $v^\mu = \frac{\d{} x^\mu}{\d{}\tau}$ satisfies $v^2 = -1$. We now generate the family of all such causal observers by exploiting the path reparameterization freedom, $\tau \to \lambda$, under which the tangent vector scales as:
\begin{equation}\label{reparametrization}
    v_\lambda^\mu = \mathscr{R}\, v^\mu \, , \qquad (v_\lambda^\mu)^2 = - \mathscr{R}^2 \, ,
\end{equation}
where $\mathscr{R} = \frac{\d{} \tau}{\d{} \lambda} > 0$. The choice of $\mathscr{R}$ physically corresponds to the local choice of the unit of the observer's time along their trajectory (see Fig.~\ref{fig:setup-1}).

The question of who formulates the second law reduces to isolating the correct trajectory and parameterization. We propose that the observer follows a  geodesic parametrized such that the expansion equals minus the non-affinity parameter:
\begin{equation}\label{non-affine-geodesic}
   \inbox{ \observer: \quad v_{\lambda} \cdot \nabla v_{\lambda} \mathrel{\hat{=}} -\theta_{v_{\lambda}} v_{\lambda}\, . }
\end{equation}
Here $\theta_{v_{\lambda}} := q_{\alpha\beta}\nabla^{\alpha}v_{\lambda}^{\beta} = \mathscr{R} \theta_{v}$. Contracting both sides of the above with $v_\lambda^\alpha$ we obtain, 
\begin{equation}\label{zeroth-law}
    v_{\lambda} \cdot \nabla (\sqrt{q}\, \mathscr{R}) \mathrel{\hat{=}} 0\, .
\end{equation}
Recalling $v_\lambda^\mu = \mathscr{R}v^\mu$ and utilizing \eqref{CD-v-inde}, \eqref{non-affine-geodesic} yields
\begin{equation}\label{geodesic-conditions}
    \kappa_v   \mathrel{\hat{=}}  0\, , \qquad a^{(v)}_\mu  \mathrel{\hat{=}} 0\, .
\end{equation}

This indicates that $v\cdot \nabla v^\mu \mathrel{\hat{=}} 0$, meaning that in the $\tau$ parameterization, the observer follows an affinely parameterized geodesic. Thus, \eqref{non-affine-geodesic} defines a non-affinely parametrized geodesic, with the parametrization defined in \eqref{zeroth-law}, which defines the ``expansion-adjusted parameterization'': the parameterization $\mathscr{R}$, which is fixed by the expansion, is chosen such that $\sqrt{q}\mathscr{R}$, the volume density on the hypersurface $\Gamma$, is conserved along the geodesic $\gamma$. For a given $v^\mu$ and surfaces $\Sigma$, any smooth orientation-preserving reparametrization \(\lambda'=f(\lambda)\), with \(f'(\lambda)>0\), gives \(dS/d\lambda'=[f'(\lambda)]^{-1}dS/d\lambda\) and therefore preserves entropy monotonicity.
%\tcr{For a fixed congruence and family of cuts, any smooth orientation-preserving reparametrization \(\lambda'=f(\lambda)\), with \(f'(\lambda)>0\), gives \(dS/d\lambda'=[f'(\lambda)]^{-1}dS/d\lambda\) and therefore preserves entropy monotonicity.}
%This physically defines an observer-dependent local thermal equilibrium along their path. 
To summarize:
\begin{tcolorbox}[
    colback=black!2, 
    colframe=black!70, 
    arc=3pt, 
    boxrule=0.5pt, 
    top=6pt, bottom=6pt, left=8pt, right=8pt,
    halign=center
]
  \itshape 
  The gravitational second law is written by a set of (locally) free-falling observers with the covariantly defined expansion-adjusted parameterization \eqref{zeroth-law}.
\end{tcolorbox}
We close this part with some comments.
\vspace*{-3mm}
\begin{enumerate}[leftmargin=2mm, itemsep=-1mm] 
    \item The above is conceptually analogous to Newtonian mechanics, where Newton's second law of mechanics is formulated by inertial observers who are defined by the first law of Newtonian mechanics.
   \item  Eqs.~\eqref{non-affine-geodesic} and \eqref{zeroth-law} are geometric and universal relations; they do not depend on the dimension $D$ and the gravity theory.\label{item-2-observer} 
    \item $\mathscr{R}$ defines the time unit as measured by the free-fall observer, which, through \eqref{zeroth-law}, is adjusted by the expansion of codimension-2 surfaces $\Sigma$.
    \item Had we instead considered a null geodesic, as in \cite{Shajiee:2026coz}, so that $\Gamma$ is a null surface, the non-affinity would be related to the surface gravity on $\Gamma$. In this case, \eqref{non-affine-geodesic} and \eqref{zeroth-law} would define a local temperature and establish a local zeroth law of thermodynamics.
  
\end{enumerate}

\begin{center}
%%%%%%%%%%%%%%%%%%%%%%%%%%%%%%%%%%%%%%%%%%%%%%%%%%%%%%%%%%%%%%%
\textbf{Derivation of the second law}\label{sec:III}
%%%%%%%%%%%%%%%%%%%%%%%%%%%%%%%%%%%%%%%%%%%%%%%%%%%%%%%%%%%%%%%
\end{center}

In our framework, the second law requires that
\begin{equation}\label{statmnt-2ndlaw-chargevar}
\delta_{\gamma}S[\lambda]\geq 0\,,     
\end{equation}
where $\delta_{\gamma}$ denotes the variation taken along the future-directed causal curve $\gamma$. Most of the second law derivations begin with the codimension-2 integral defining the entropy, such as \eqref{entropy-def-EH}, and seek to establish \eqref{statmnt-2ndlaw-chargevar}. We instead adopt a somewhat different approach.

By applying Stokes’ theorem, we recast the entropy variation as a codimension-1 integral over $\Gamma$:
{\begin{equation}\label{2nd-law-chargvar-stokes}
    \begin{split}
   \hspace*{-2.5mm}     \delta_{\gamma}S[\lambda] & = \frac{1}{4G\hbar}\int_{\Sigma(\lambda)} \d{}^{D-2}x\, {\cal L}_{v_\lambda}{\sqrt{q}}  \\
        & =  \frac{1}{4G\hbar}\left[\int_{\Sigma(\lambda)} \d{}^{D-2}x{\cal L}_{v_\lambda}{\sqrt{q}} - \int_{\Sigma(\infty)}\d{}^{D-2}x{\cal L}_{v_\lambda}{\sqrt{q}}\right] \\
        & = - \frac{1}{4G\hbar}\int_{\Gamma} \d{}^{D-2}x\, \d{}\lambda\, \ {\cal L}_{v_\lambda}(  {\cal L}_{v_\lambda}{\sqrt{q}})\, ,
    \end{split}
\end{equation}
where ${\cal L}_{v_\lambda}$ is the Lie derivative along the non-affinely parameterized geodesic with velocity vector $v_\lambda^\mu = \mathscr{R}v^\mu$. In the second line, in accord with other works on the second law literature, e.g., \cite{Wall:2015raa, Hollands:2022fkn}, we have reasonably assumed that, as depicted in Fig~\ref{fig:setup-2}, the system settles into a non-expanding ``equilibrium state'', ${\cal L}_{v_\lambda}\sqrt{q}|_{\lambda\to \infty}=0$, which together with \eqref{zeroth-law} means we take $\mathscr{R}=1$ at large $\lambda$. We emphasize that in \eqref{2nd-law-chargvar-stokes} the integral is taken over the timelike causal hypersurface $\Gamma$, rather than over a full or partial spacelike Cauchy surface $\mathcal{C}$.

To evaluate the variation $\delta_{\gamma}S[\lambda]$, we apply the observer conditions proposed above \eqref{non-affine-geodesic} and \eqref{zeroth-law}. Using the causal Raychaudhuri equation for the affinely parametrized geodesic congruence $v^\mu$ ($v \cdot \nabla v = 0$), 
\begin{equation}\label{Raych-Eqs}
    v^{\mu}\nabla_{\mu} \theta_v \mathrel{\hat{=}} - \frac{\theta_v^2}{D-2} - 2\omega_\mu \omega^\mu - N_{v}^2 - R_{vv}\, .
\end{equation}
Note that the Raychaudhuri equation contains the term $\nabla_{\mu}a^{\mu}$. Using $a_\mu\mathrel{\hat{=}}0$, one can readily show that $\nabla_{\mu}a^{\mu} \mathrel{\hat{=}}0$ as well. Substituting \eqref{Raych-Eqs} into \eqref{2nd-law-chargvar-stokes} alongside \eqref{zeroth-law} and after straightforward algebra, we arrive at our main result (see \cite{Shajiee:2025cxl} for details of the derivation):
\begin{equation}\label{delta-S-main}
\inbox{
  \begin{aligned}
    \delta_{\gamma}S[\tau] 
    = \frac{1}{4G\hbar}\int_{\Gamma} \mathrm{d}^{D-2}x\, \mathrm{d}\tau\, \sqrt{q}\, \mathscr{R} \times {} & \\
   \bigg[ \frac{\theta_{v}^{2}}{D-2} + 2\omega_\mu \omega^\mu + N_v^2 + R_{vv} \bigg]\, . &
  \end{aligned}
}   
\end{equation}
Here, $N_v^2 = N^{(v)}_{\mu\nu}N_{(v)}^{\mu\nu}$ and $R_{vv} = R_{\mu\nu} v^\mu v^\nu$. 

The squared expansion $\theta_v^2$, the rotational term $\omega_\mu \omega^\mu$, and the shear $N_v^2$ are all manifestly positive-definite. Thus, only the sign of the Ricci curvature projection $R_{vv}$ remains undetermined. Recalling the Einstein field equations, $R_{\mu\nu}= 8\pi G (T_{\mu\nu}-\frac{T}{D-2} g_{\mu\nu})$, we find
\begin{equation}\label{Rvv-SEC}
    R_{vv} = 8\pi G \Big(T_{\mu\nu}-\frac{T}{D-2} g_{\mu\nu} \Big) v^\mu v^\nu\, . 
\end{equation} 
Thus, the Strong Energy Condition (SEC) on the matter sector \cite{Ford:1994bj, Curiel:2014zba, Wald:1984rg, Kontou:2020bta} guarantees the condition $R_{vv}\geq 0$. Strictly speaking, our proof only requires the ``integrated SEC'' along segments of the future-oriented causal curve $\gamma$.

We close this part with some comments on \eqref{2nd-law-chargvar-stokes}.
\vspace*{-3mm}
\begin{enumerate}[leftmargin=1.5mm, itemsep=-1mm] 
\item  $\delta_{\gamma}S[\lambda]$ is the time derivative of the entropy and hence has dimension of energy in natural units. The Right-Hand-Side, too, has dimension of energy. That is, \eqref{2nd-law-chargvar-stokes} is basically an energy-balance law written for the free-fall observer on an expansion-adjusted parametrized geodesic.
\item  The integral measure $\sqrt{q}\mathscr{R}$ is a constant along the observer path $\gamma$, \eqref{zeroth-law}. 
\item  The $\theta_v^2$ and $\omega^2$ terms are observer-dependent and could have been absorbed into the observer's definition \cite{Shajiee:2025cxl}. If we did so, we would have lost the universality of the observer (cf. item \ref{item-2-observer} in the last paragraph of the previous part). 
\item  The definition of the entropy \eqref{entropy-general-def} and hence \eqref{2nd-law-chargvar-stokes} apply to generic dynamical $\Gamma$ or $\Sigma$; they need not be close to stationarity or be a (Killing) horizon. In this sense, they are similar to the ``dynamical entropy'' introduced and discussed in \cite{Shajiee:2026coz}.
\item Our analyses and results here are applicable to any causal, timelike, and null observer. For the specific case of null observers, one may take the null limit of a generic timelike observer. The details of such a limit are presented in \cite{Shajiee:2025cxl}. Here we quote the results: if $l^\mu$ is along the velocity vector of the null observer, and $\gamma$ is a null geodesic with affine parameter $\tau$, then
\begin{equation}\label{Null-2nd-law}
    \delta_{\gamma}S[\tau] 
    = \frac{1}{4G\hbar}\int_{\Gamma} \mathrm{d}^{D-2}x\, \mathrm{d}\tau\, \sqrt{q}\,\tilde{\mathscr{R}} \bigg[ \frac{\theta_{l}^{2}}{D-2} + N_l^2 + R_{ll} \bigg]\, ,
\end{equation}
\end{enumerate}
where $\tilde{\mathscr{R}}$ is the non-affinity parameter, as in the timelike case. 
The second law is guaranteed when the integral of $R_{ll}$ over the null curve is non-negative, which, for Einstein gravity, implies the integrated null energy condition for the matter fields, in accord with the standard results in the literature \cite{Wall:2009wm, Bousso:2015eda}.

\begin{center}
%%%%%%%%%%%%%%%%%%%%%%%%%%%%%%%%%%%%%%%%%%%%%
\textbf{Discussion and outlook }\label{sec:V}
%%%%%%%%%%%%%%%%%%%%%%%%%%%%%%%%%%%%%%%%%%%%%
\end{center}

Since Unruh's seminal paper \cite{Unruh:1976db}, it is known that the thermodynamic description of gravitating systems depends on the observer; however, this fact is not noted and discussed as much as it deserves. Temperature (cf. Unruh's work \cite{Unruh:1976db}), other chemical potentials, and other charges like mass and angular momentum, especially in the context of black hole thermodynamics, are well-known to be observer-dependent \cite{Grumiller:2022qhx}; the entropy, however, is a (Noether) surface charge \cite{Wald:1993nt, Shajiee:2026coz}  which is not expected to be observer-dependent. Here, we explored the observer-dependence of the entropy and the second law. 

There are some different proposals to define the entropy in gravitational systems; here we focused on the one introduced in \cite{Shajiee:2025cxl}, see also \cite{Jacobson:2015uqa, Godazgar:2020kqd}, as the surface charge associated with local boosts. This definition does not rely on any vector field (as a symmetry generator) and is hence an observer-independent notion, in the sense of the usual metric formulation of gravity. 

Nonetheless, to discuss the second law, there is no escape from considering observers. Our proposal starts with free-falling observers who follow causal (timelike or null) geodesics. The second law is about recording changes in the entropy as time advances, and recalling that on a geodesic one still has the freedom to choose the path parametrization, the local unit of time along the path, one can ask if the path parametrization can affect the derivation/proof of the second law. The answer is expected to be yes, because the change in the entropy involves terms that are observer-dependent (expansion-squared terms), which in particular depend on the adapted local time unit. These considerations, and the universality of the second law and the observer who records it, led us to the particular, purely geometric, non-affine ``expansion-adjusted parametrization'' \eqref{non-affine-geodesic}. As we showed, cf. \eqref{delta-S-main}, for this class of observers the second law is guaranteed upon integrated-SEC.

All ingredients in our analyses, the entropy, entropy variation, and the observer choice, are (quasi)-local and covariant. Moreover, they do not rely on Killing horizons, stationarity, or perturbations around equilibrium. Our constructions apply to any gravitational system, extending the existing literature on the notion of entropy, e.g., in \cite{Wald:1993nt, Iyer:1994ys, Hollands:2022fkn, Hollands:2024vbe, Ashtekar:2002ag, Ashtekar:2003hk, Wall:2009wm, Wall:2011hj} and on the derivation of the second law for timelike (as well as null) observers. 

While we showcased the analyses for Einstein gravity theory, our construction is expected to generalize to any diffeomorphism-invariant gravity theory, which we will present in a separate, more detailed publication. We close with a comment that the integrated-SEC condition may be violated in accelerating cosmologies, e.g., the late dark energy-dominated universe or early universe during inflation. Studying the implications of our derivation and analyses for these cases is an interesting subject for future studies.

\vskip 2mm
%\begin{center}
%%%%%%%%%%%%%%%%%%%%%%%%%%%%%%%%%%%%%%%%%%%%%
\textbf{Acknowledgment.} %\label{sec:V}
%%%%%%%%%%%%%%%%%%%%%%%%%%%%%%%%%%%%%%%%%%%%%
%\end{center}
%%%%%%%%%%%%%%%%%%%%%%%%%%%%%%%%%%%%%%%%%%%%%%%%%
%\section*{Acknowledgment}\label{Acknow}
%%%%%%%%%%%%%%%%%%%%%%%%%%%%%%%%%%%%%%%%%%%%%%%%%
The work of VT acknowledges Iran National Science Foundation (INSF) grant No. 40506209. We acknowledge the support through INSF research chair No. 4045163. 

%$\vspace*{-1mm}$
%%%%%%%%%%%%%%%%%%%%%%%%%%%%%%%%%%%%%%%%%%%%%%%%%%%%
\appendix
%%%%%%%%%%%%%%%%%%%%%%%%%%%%%%%%%%%%%%%%%%%%%%%%%%%%

$\vspace*{-7mm}$
%\vskip -5mm

%%%%%%%%%%%%%%%%%%%%%%%%%%%%%%%%%%%%%%%%%%%%%%%%%%%%

\bibliographystyle{fullsort.bst}
\bibliography{reference}
\end{document}